\documentclass[3p,times,twocolumn]{elsarticle}

\usepackage{ecrc}

\volume{00}

\firstpage{1}

\journalname{Nuclear Physics B Proceedings Supplement}

\runauth{}

\jid{nuphbp}

\jnltitlelogo{Nuclear Physics B Proceedings Supplement}

\usepackage{amssymb}
\usepackage[figuresright]{rotating}

\def\nin{\noindent}
\def\beq{\begin{equation}}
\def\eeq{\end{equation}}
\def\bea{\begin{eqnarray}}
\def\eea{\end{eqnarray}}

\begin{document}

\begin{frontmatter}




\title{{\large Lightest pseudoscalar exchange contribution to light-by-light scattering piece of the muon $g-2$}}

 \author[label1]{Pablo Roig\corref{cor1}}
  \address[label1]{Instituto de F\'{\i}sica, Universidad Nacional Aut\'onoma de M\'exico,
Apartado Postal 20-364, 01000 M\'exico D.F., M\'exico.}
%\ead{pabloroig@fisica.unam.mx}

 \author[label2]{Adolfo Guevara}
  \address[label2]{Departamento de F\'isica, Centro de Investigaci\'on y de Estudios Avanzados, Apartado Postal 14-740, 07000 M\'exico D.F., M\'exico.}
%\ead{aguevara@fis.cinvestav.mx}

\author[label2]{Gabriel L\'opez Castro}
%\ead{glopez@fis.cinvestav.mx}

\begin{abstract}
%\noindent
Lightest pseudoscalar ($P=$ $\pi^0$, $\eta$, $\eta'$) exchange contribution to the light-by-light ($LbL$) scattering piece of the muon anomaly, $a_\mu=(g_\mu-2)/2$, has been 
evaluated using a resonance chiral Lagrangian ($R\chi L$). Best description of pion transition form-factor ($TFF$) data is obtained with only tiny violations of one of the 
relations in the minimal consistent set of short-distance constraints on the anomalous $R\chi L$ couplings. $\eta^{(')}$ $TFF$, predicted in terms of the $\pi$ $TFF$ and the 
$\eta-\eta^{'}$ mixing, are in good agreement with measurements. With this input, we obtain $a_\mu^{P,LbL}\,=\,(10.47\pm0.54)\cdot 10^{-10}$, consistent with the reference 
determinations in the literature, albeit with smaller error.
\end{abstract}


\end{frontmatter}

% \linenumbers

%\section{}
%\label{}

\begin{figure}[h!]
\begin{center}
%\vspace*{1.25cm}
\includegraphics[scale=0.3]{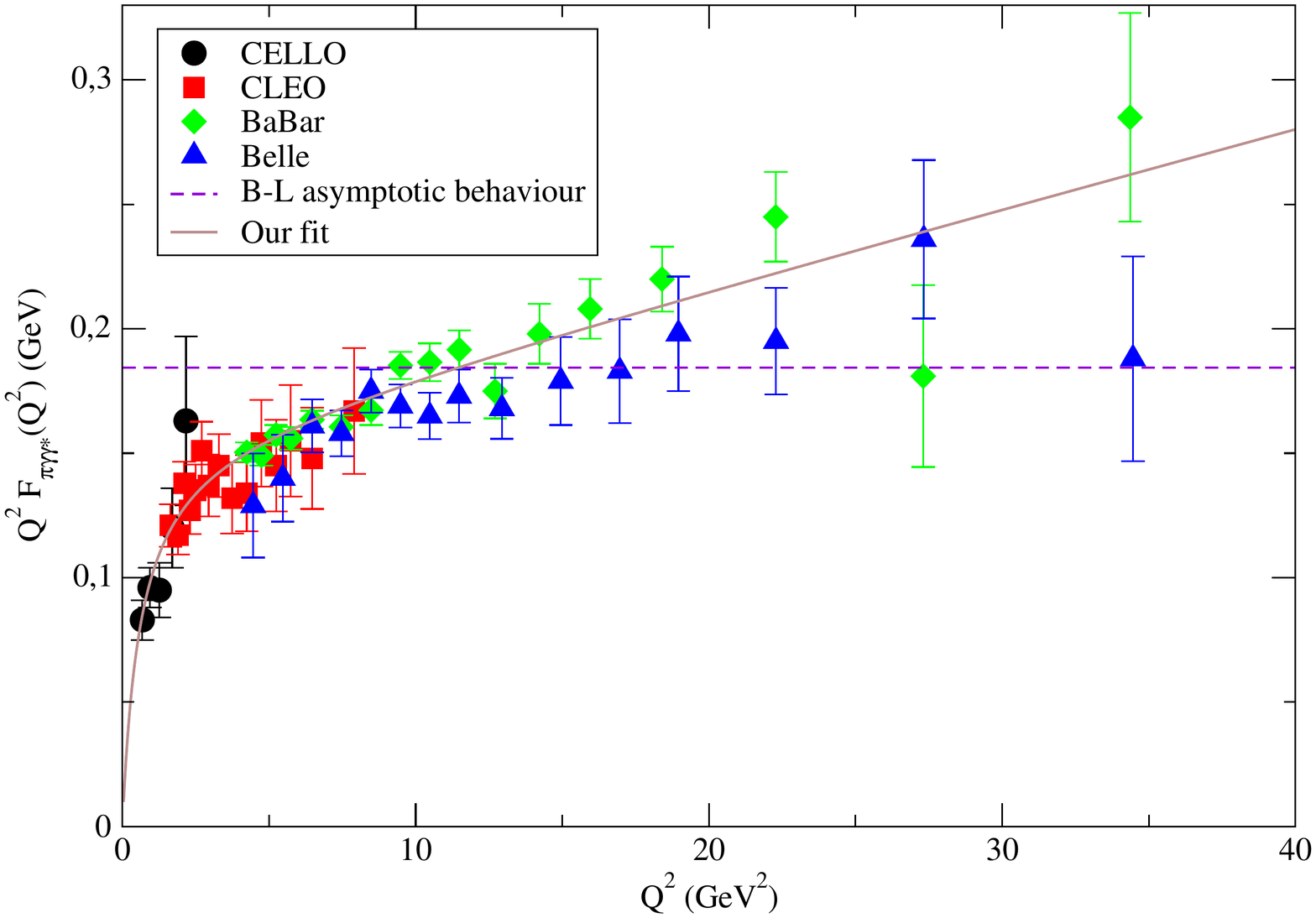}
\caption[]{\label{fig:piggFF} \small{CELLO \cite{Behrend:1990sr}, CLEO \cite{Gronberg:1997fj}, BaBar \cite{Aubert:2009mc} and Belle \cite{Uehara:2012ag} data for the $\pi$TFF 
are confronted to our best fit result using the form-factor in eq.~(\ref{fit piggFF}).}}
\end{center}
\end{figure}

\nin
%%%%%%%%%%%%
%\hspace*{0.1cm}
\indent The anomalous magnetic dipole moment of the muon, $a_\mu$, is one the most precisely measured \cite{BNL} and accurately predicted \cite{Theo} observables. Since 2000, it exhibits 
a persistent discrepancy in the ballpark of three standard deviations, $a_\mu^{\mathrm{exp}}-a_\mu^{\mathrm{th}}\,=\,(29\pm9)\cdot10^{-10}$, $(3.3\sigma)$~\cite{Theo}.\\
\indent As an electromagnetic property $a_\mu$ is, first, a stringent test of $QED$. The one loop contribution computed by Schwinger \cite{Schwinger:1948iu} fixes its size universally 
to $\sim \alpha/(2\pi)\sim10^{-3}$ for all charged leptons, which differs by six orders of magnitude with the current discrepancy $\sim3\cdot10^{-9}$. 
A tremendous effort in computing higher-order terms lead to the complete five-loop $QED$ result \cite{Aoyama:2012wk}. Its error is four orders of magnitude smaller than the previous value 
and the $\mathcal{O}(\alpha^5)$ term is only $\sim5\cdot10^{-11}$. Therefore, this anomaly cannot be attributed to uncalculated or imprecise $QED$ contributions.\\
\indent Conversely, hadronic effects are important in $a_\mu$ \footnote{Electroweak effects are small and well under control at the required precision, $a_\mu^{EW}\,=\,(153.6\pm1.0)
\cdot10^{-11}$ \cite{EW}.}: hadronic vacuum polarization ($LO+NLO$) contributes $(680.7\pm4.7)\cdot10^{-10}$ to $a_\mu$ \cite{HVP}. 
Although with much smaller central value, the Hadronic $LbL$ contribution ($HLbL$) is crucial for the final theoretical uncertainty: 
$a_\mu^{HLbL}\,=\,(11.6\pm4.0)\cdot10^{-10}$ \cite{Theo}, if we stick to the most conservative estimate.\\
\indent The total error on the Standard Model prediction for $a_\mu$, $6.3\cdot10^{-10}$, nearly equals the current experimental uncertainty, $6.4\cdot10^{-10}$. However, forthcoming 
experiments at FNAL and J-PARC \cite{NewExp} will soon reduce the latter to a fourth. This urges theoreticians to achieve a similar error reduction to benefit fully from the 
precision of these measurements. This is our main motivation to revisit the dominant contribution to $a_\mu^{HLbL}$ given by $P$ exchange.\\
\indent We have reconsidered \cite{Us} the lightest pseudoscalar ($P=$ $\pi^0$, $\eta$, $\eta'$) exchange contribution to $a_\mu^{HLbL}$, which is one of the possible intermediate 
states in the four-point $VVVV$ Green function with one real and three virtual photons that enters $a_\mu^{HLbL}$. This contribution turns out to basically saturate it, due 
to approximate cancellations between the remaining pieces \cite{Theo}. Despite a simultaneous chiral and large-$N_C$ expansion has been suggested \cite{deRafael:1993za} 
to tackle $a_\mu^{HLbL}$, the relative size of the various terms is not fully understood yet. This raises reasonable doubts on the errors obtained studying invividual 
contributions isolately, which might translate into an increased overall error for $a_\mu^{HLbL}$ \cite{Nyffeler:2013lia}.\\
\indent The main difficulty for computing $a_\mu^{HLbL}$ has been that, contrary to the hadronic vacuum polarization, there was no way to relate it to measurements using dispersion 
relations and the optical theorem. However, very recently a formalism has been put forward \cite{Colangelo:2014dfa}, which would allow to extract the dominant one- and 
two-pion exchange contributions directly from data.\\
\indent At the time being, nevertheless, the most precise experimental information that can be obtained on $a_\mu^{P,HLbL}$ comes from the corresponding $TFF$, where one photon is 
real to a very good approximation and the form-factor is measured as a function of the virtuality of the other photon up to roughly $6$ GeV. It has been shown \cite{Theo} that 
demanding an appropriate short-distance behaviour \cite{OPE} to the $P$ $TFF$ and to the related $VVP$ Green's function turns out to be crucial for the reliability of the 
$a_\mu^{P,HLbL}$ value. Dedicated studies of this question \cite{VFRO} have found that using $R\chi L$ \cite{RChT} (rooted in the large-$N_C$ limit of QCD \cite{Nc} and 
chiral symmetry \cite{ChPT}) there exists a consistent set of short-distance constraints on the $VVP$ Green function 
and related form factors \cite{Roig:2013baa}. For this, the antisymmetric tensor formalism needs to be employed \cite{RuizFemenia:2003hm} and pseudoscalar resonances be active 
degrees of freedom \cite{Kampf:2011ty}. However, it is still an open question whether all short-distance constraints are already imposed working this way, or there are new genuine 
relations arising from the $VVVV$ Green function which have not been considered yet \cite{SDConstr}.\\
\indent In this framework the $\pi$ $TFF$ can be written \cite{Roig:2013baa, Kampf:2011ty}
\begin{equation}\label{fit piggFF}
 \mathcal{F}_{\pi^0\gamma\gamma}(Q^2)\,=\,-\frac{F}{3}\frac{Q^2\left(1+32\sqrt{2}\frac{P_2F_V}{F^2}\right)+\frac{N_C}{4\pi^2}\frac{M_V^4}{F^2}}{M_V^2(M_V^2+Q^2)}\,,\;\;
\end{equation}
and one of the high-energy constraints in the minimal consistent set demands that $P_2$ cancels the $\mathcal{O}(Q^0)$ term for $Q^2\to\infty$. We find that allowing a $4\%$ 
violation of this condition yields the best fit to current data \footnote{This violation is not due to the difference between BaBar and Belle data points. Excluding BaBar 
data the violation is only reduced to $3\%$.}. Fig.~\ref{fig:piggFF} compares our best fit result to all available data.

\begin{figure}[h!]
\begin{center}
%\vspace*{1.25cm}
\includegraphics[scale=0.3]{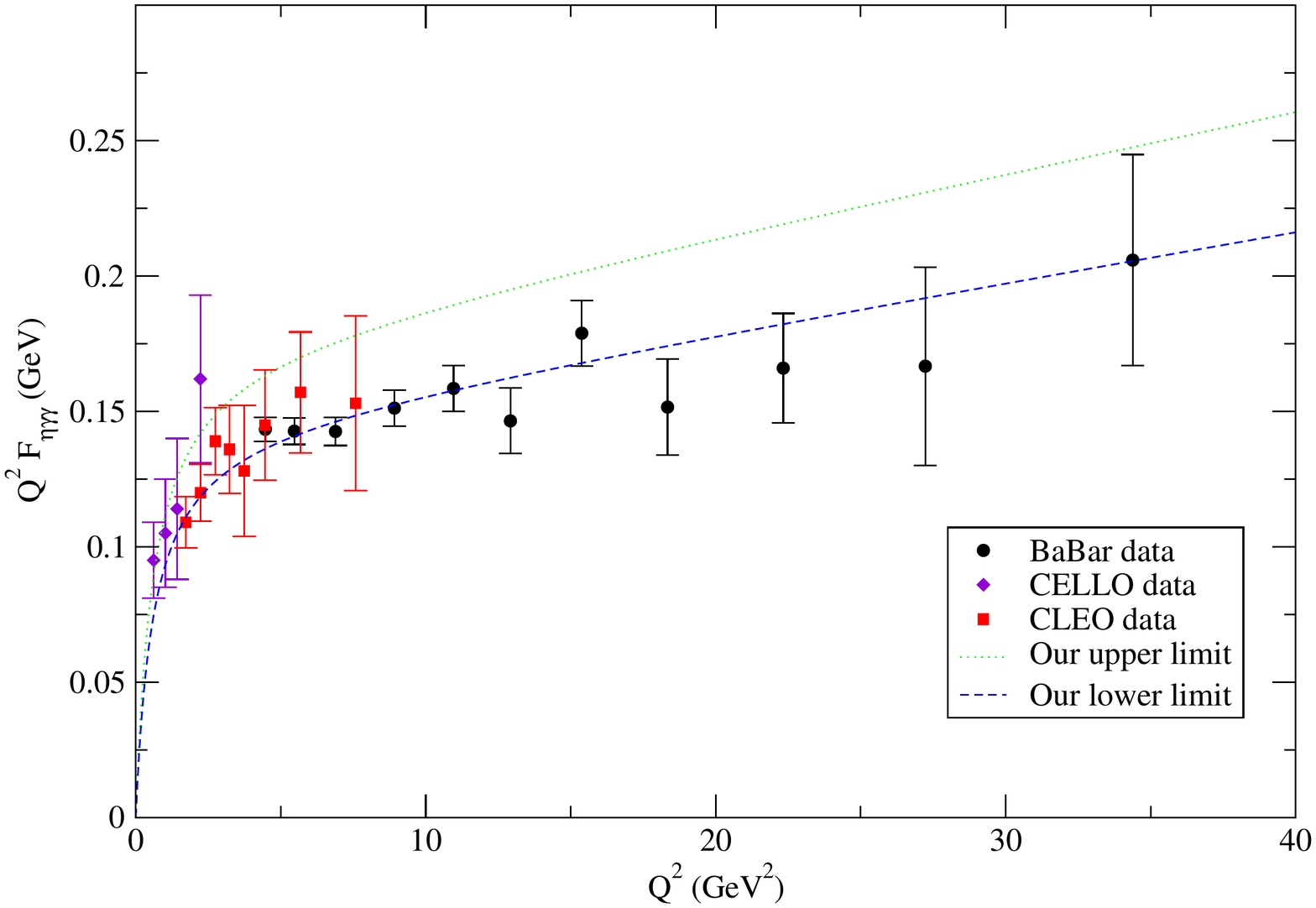}
\caption[]{\label{fig:etaTFF} \small{Our predictions for the $\eta$TFF using the $\pi$ TFF (\ref{fit piggFF}) and the 
$\eta$-$\eta^\prime$ mixing are confronted to BaBar~\cite{BABAR:2011ad}, CELLO~\cite{Behrend:1990sr} and CLEO~\cite{Gronberg:1997fj} data.}}
\end{center}
\end{figure}

\begin{figure}[h!]
\begin{center}
%\vspace*{1.25cm}
\includegraphics[scale=0.3]{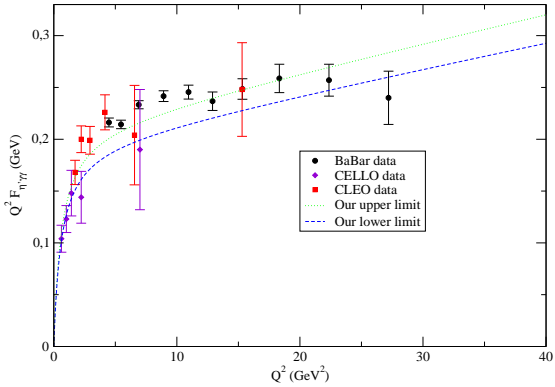}
\caption[]{\label{fig:etapTFF} \small{Our predictions for the $\eta^\prime$TFF using the $\pi$ TFF (\ref{fit piggFF}) and the 
$\eta$-$\eta^\prime$ mixing are confronted to BaBar~\cite{BABAR:2011ad}, CELLO~\cite{Behrend:1990sr} and CLEO~\cite{Gronberg:1997fj} data.}}
\end{center}
\end{figure}

\indent The fully off-shell form factor is also needed to evaluate $a_{\mu}^{P,HLbL}$ \cite{Theo}. It depends only on one additional coupling unrestricted by high-energy behaviour, which 
can be fixed analysing the $\pi(1300)\to\gamma\gamma$ and $\pi(1300)\to\rho\gamma$ decays \cite{Kampf:2011ty}. Upon the required integrations \cite{Theo} one finds

\begin{equation}\label{Result virtual pion}
 a_{\mu}^{\pi^0,HLbL}\,=\,\left(6.66\pm0.21\right)\cdot 10^{-10}\,,
\end{equation}
where the corresponding contributions to the error are discussed in detail in our paper \cite{Us}.\\
\indent Chiral dynamics and the $\eta$-$\eta^\prime$ mixing (whose uncertainty saturates the error on $a_{\mu}^{\eta^{(\prime)},HLbL}$) allow to relate the $\pi$ $TFF$ to the 
$\eta^{(\prime)}$ $TFF$ \cite{Us}. Our predictions are compared to data in Figs.~\ref{fig:etaTFF} and \ref{fig:etapTFF}.

\indent The corresponding contributions to  $a_{\mu}^{P,HLbL}$ being

\begin{equation}\label{Result virtual eta}
 a_{\mu}^{\eta,HLbL}\,=\,\left(2.04\pm0.44\right)\cdot 10^{-10}\,,
\end{equation}

\begin{equation}\label{Result virtual eta '}
a_{\mu}^{\eta^\prime,HLbL}\,=\,\left(1.77\pm0.23\right)\cdot 10^{-10}\,.
\end{equation}

\indent Our main result is the value 
\begin{equation}
a_{\mu}^{P,HLbL}\,=\,\left(10.47\pm0.54\right)\cdot 10^{-10}\ , 
\end{equation}
for the contribution of the three lightest pseudoscalar 
mesons ($\pi^0$, $\eta$ and $\eta^\prime$) to the muon anomaly, 
which is in good agreement with the two reference values: $\left(9.9\pm1.6\right)\cdot 10^{-10}$ (Jegerlehner and Nyffeler) and 
$\left(11.4\pm1.3\right)\cdot 10^{-10}$ (Prades, de Rafael and Vainshtein) \cite{Theo} but has a reduced error, mainly thanks to the 
new BaBar and Belle data on the $P$ $TFF$ extending to larger energies.\\
\indent Using the values in the literature for the remaining contributions to $a_\mu^{HLbL}$ \cite{Theo} yields
\begin{equation}
a_{\mu}^{HLbL}\,=\,\left(11.8\pm2.0\right)\cdot 10^{-10}\ ,
\end{equation}
which would translate into a theoretical uncertainty on $a_\mu$ of $\pm5.1\cdot10^{-10}$, $\sim20\%$ smaller than the current estimate.\\
\indent Finally, we also propose \cite{Us} the measurement of $\sigma(e^+e^-\to\pi^0\mu^+\mu^-)$ and the corresponding differential distribution as a function of the di-muon invariant mass to 
better characterize the $P$ $TFF$ and hopefully further reduce the error of $a_{\mu}^{P,HLbL}$.
%%%%%%%%%%%%%%%%%%%%%%%%%%%
\section*{Acknowledgements}
%\nin
%%%%%%%%%%%%%%%%
P.~R.~ acknowledges receiving an ICHEP14 grant covering his participation. Discussions on this topic with B.~Malaescu and H.~Hayashii during ICHEP14 are very much 
appreciated. This work is partly funded by the Mexican Government through CONACYT and DGAPA (PAPIIT IN106913).



%% References

%\nocite{*}
%\bibliographystyle{elsarticle-num}
%\bibliography{martin}

\begin{thebibliography}{00}



\bibitem{BNL}
%\bibitem{Bennett:2006fi}
  G.~W.~Bennett {\it et al.}  [Muon G-2 Collaboration],
  %``Final Report of the Muon E821 Anomalous Magnetic Moment Measurement at BNL,''
  Phys.\ Rev.\ D {\bf 73} (2006) 072003;
%  [hep-ex/0602035].
  %%CITATION = HEP-EX/0602035;%%
%\bibitem{Bennett:2004pv}
%  G.~W.~Bennett {\it et al.}  [Muon g-2 Collaboration],
  %``Measurement of the negative muon anomalous magnetic moment to 0.7 ppm,''
  Phys.\ Rev.\ Lett.\  {\bf 92} (2004) 161802.
%  [hep-ex/0401008].
  %%CITATION = HEP-EX/0401008;%%

\bibitem{Theo}
  F.~Jegerlehner and A.~Nyffeler,
  %``The Muon g-2,''
  Phys.\ Rept.\  {\bf 477} (2009) 1.
%  [arXiv:0902.3360 [hep-ph]].
  %%CITATION = ARXIV:0902.3360;%%
%\bibitem{Prades:2009tw}
  J.~Prades, E.~de Rafael and A.~Vainshtein,
  %``Hadronic Light-by-Light Scattering Contribution to the Muon Anomalous Magnetic Moment,''
  Advanced series on directions in high energy physics. Vol. 20.
%  [arXiv:0901.0306 [hep-ph]].
  %%CITATION = ARXIV:0901.0306;%%

\bibitem{Schwinger:1948iu}
  J.~S.~Schwinger,
  %``On Quantum electrodynamics and the magnetic moment of the electron,''
  Phys.\ Rev.\  {\bf 73} (1948) 416.
  %%CITATION = PHRVA,73,416;%%

\bibitem{Aoyama:2012wk}
  T.~Aoyama, M.~Hayakawa, T.~Kinoshita and M.~Nio,
  %``Complete Tenth-Order QED Contribution to the Muon g-2,''
  Phys.\ Rev.\ Lett.\  {\bf 109} (2012) 111808.
%  [arXiv:1205.5370 [hep-ph]].
  %%CITATION = ARXIV:1205.5370;%%

\bibitem{EW}
%\bibitem{Czarnecki:1995sz}
  A.~Czarnecki, B.~Krause and W.~J.~Marciano,
  %``Electroweak corrections to the muon anomalous magnetic moment,''
  Phys.\ Rev.\ Lett.\  {\bf 76} (1996) 3267.
%  [hep-ph/9512369].
  %%CITATION = HEP-PH/9512369;%%
%\bibitem{Knecht:2002hr}
  M.~Knecht, S.~Peris, M.~Perrottet and E.~De Rafael,
  %``Electroweak hadronic contributions to the muon (g-2),''
  JHEP {\bf 0211} (2002) 003.
%  [hep-ph/0205102].
  %%CITATION = HEP-PH/0205102;%%
%\bibitem{Czarnecki:2002nt}
  A.~Czarnecki, W.~J.~Marciano and A.~Vainshtein,
  %``Refinements in electroweak contributions to the muon anomalous magnetic moment,''
  Phys.\ Rev.\ D {\bf 67} (2003) 073006
   [Erratum-ibid.\ D {\bf 73} (2006) 119901].
%  [hep-ph/0212229].
  %%CITATION = HEP-PH/0212229;%%
%\bibitem{Gnendiger:2013pva}
  C.~Gnendiger, D.~St\"ockinger and H.~St\"ockinger-Kim,
  %``The electroweak contributions to $(g-2)_\mu$ after the Higgs boson mass measurement,''
  Phys.\ Rev.\ D {\bf 88} (2013) 5,  053005.
%  [arXiv:1306.5546 [hep-ph]].
  %%CITATION = ARXIV:1306.5546;%%
  
\bibitem{HVP}
%\bibitem{Davier:2010nc}
  M.~Davier, A.~Hoecker, B.~Malaescu and Z.~Zhang,
  %``Reevaluation of the Hadronic Contributions to the Muon g-2 and to alpha(MZ),''
  Eur.\ Phys.\ J.\ C {\bf 71} (2011) 1515
   [Erratum-ibid.\ C {\bf 72} (2012) 1874].
%  [arXiv:1010.4180 [hep-ph]].
  %%CITATION = ARXIV:1010.4180;%%
%\bibitem{Krause:1996rf}
  B.~Krause,
  %``Higher order hadronic contributions to the anomalous magnetic moment of leptons,''
  Phys.\ Lett.\ B {\bf 390} (1997) 392.
%  [hep-ph/9607259].
  %%CITATION = HEP-PH/9607259;%%

\bibitem{NewExp}
%\bibitem{Benayoun:2014tra}
%  M.~Benayoun, J.~Bijnens, T.~Blum, I.~Caprini, G.~Colangelo, H.~Czyż, A.~Denig and C.~A.~Dominguez {\it et al.},
 T.~Mibe contribution to  
% ``Hadronic contributions to the muon anomalous magnetic moment Workshop. $(g-2)_{\mu}$: Quo vadis? Workshop. Mini proceedings,''
  arXiv:1407.4021 [hep-ph].
  %%CITATION = ARXIV:1407.4021;%%
 G.~Venanzoni for the new MUON g-2 Coll., these proceedings.

\bibitem{Us}
%\bibitem{Roig:2014uja}
  P.~Roig, A.~Guevara and G.~L\'opez Castro,
  %``The VV'P form factors in resonance chiral theory and the pi-eta-eta' light-by-light contribution to the muon g-2,''
  Phys.\ Rev.\ D {\bf 89} (2014) 073016.
%  [arXiv:1401.4099 [hep-ph]].
  %%CITATION = ARXIV:1401.4099;%%

\bibitem{deRafael:1993za}
  E.~de Rafael,
  %``Hadronic contributions to the muon g-2 and low-energy QCD,''
  Phys.\ Lett.\ B {\bf 322} (1994) 239.
%  [hep-ph/9311316].
  %%CITATION = HEP-PH/9311316;%%

\bibitem{Nyffeler:2013lia}
  A.~Nyffeler,
  %``Status of hadronic light-by-light scattering in the muon $g-2$,''
  Nuovo Cim.\ C {\bf 037} (2014) 02,  173.
%  [arXiv:1312.4804 [hep-ph]].
  %%CITATION = ARXIV:1312.4804;%%

\bibitem{Colangelo:2014dfa}
  G.~Colangelo, M.~Hoferichter, M.~Procura and P.~Stoffer,
  %``Dispersive approach to hadronic light-by-light scattering,''
  arXiv:1402.7081 [hep-ph].
  %%CITATION = ARXIV:1402.7081;%%

\bibitem{OPE}
  M. A. Shifman, A. I. Vainshtein, and V. I. Zakharov,
%%%QCD and Resonance Physics: Sum Rules,
  Nucl. Phys. B{\bf 147} (1979) 385;       %%%–447.
%
%%%M. A. Shifman, A. I. Vainshtein, and V. I. Zakharov,
%%%QCD and Resonance Physics: Applications,
%%%    Nucl. Phys.  {\bf 147} (1979)
 448.       %%%–518. B

\bibitem{VFRO}
%\bibitem{Moussallam:1994xp}
  B.~Moussallam,
  %``Chiral sum rules for parameters of the order six Lagrangian in the W-Z sector and application to pi0, eta, eta-prime decays,''
  Phys.\ Rev.\ D {\bf 51} (1995) 4939;
%  [arXiv:hep-ph/9407402];
  %%CITATION = HEP-PH/9407402;%%
%\bibitem{Moussallam:1997xx}
%  B.~Moussallam,
  %``A Sum rule approach to the violation of Dashen's theorem,''
  Nucl.\ Phys.\ B {\bf 504} (1997) 381.
%  [arXiv:hep-ph/9701400].
  %%CITATION = HEP-PH/9701400;%%
%%%QCD short distance constraints and hadronic approximations
    J. Bijnens, E. Gamiz, E. Lipartia and J. Prades,
    JHEP {\bf 0304} (2003) 055.
%    [arXiv:hep-ph/0304222].
%\bibitem{Knecht:2001xc}
  M.~Knecht and A.~Nyffeler,
  %``Resonance estimates of O(p**6) low-energy constants and QCD short distance constraints,''
  Eur.\ Phys.\ J.\ C {\bf 21} (2001) 659.
%  [arXiv:hep-ph/0106034].
  %%CITATION = HEP-PH/0106034;%%

\bibitem{RChT}
% \bibitem{Ecker:1988te}
  G.~Ecker \textit{et. al.}, %J.~Gasser, A.~Pich and E.~de Rafael,
  %``The Role of Resonances in Chiral Perturbation Theory,''
  Nucl.\ Phys.\ B {\bf 321} (1989) 311,
  %%CITATION = NUPHA,B321,311;%%
%\bibitem{Ecker:1989yg}
%  G.~Ecker, J.~Gasser, H.~Leutwyler, A.~Pich and E.~de Rafael,
  %``Chiral Lagrangians for Massive Spin 1 Fields,''
  Phys.\ Lett.\ B {\bf 223} (1989) 425.
  %%CITATION = PHLTA,B223,425;%%

\bibitem{Nc}
%\bibitem{'tHooft:1973jz}
  G.~'t Hooft,
  %``A Planar Diagram Theory for Strong Interactions,''
  Nucl.\ Phys.\ B {\bf 72} (1974) 461;
  %%CITATION = NUPHA,B72,461;%%
%\bibitem{'tHooft:1974hx}
%  G.~'t Hooft,
  %``A Two-Dimensional Model for Mesons,''
%  Nucl.\ Phys.\ B
 {\bf 75} (1974) 461.
  %%CITATION = NUPHA,B75,461;%%
%\bibitem{Witten:1979kh}
  E.~Witten,
  %``Baryons in the 1/n Expansion,''
  Nucl.\ Phys.\ B {\bf 160} (1979) 57.
  %%CITATION = NUPHA,B160,57;%%

 \bibitem{ChPT}
%\bibitem{Weinberg:1978kz}
  S.~Weinberg,
  %``Phenomenological Lagrangians,''
  Physica A {\bf 96} (1979) 327.
  %%CITATION = PHYSA,A96,327;%%
%\bibitem{Gasser:1983yg}
  J.~Gasser and H.~Leutwyler,
  %``Chiral Perturbation Theory to One Loop,''
  Annals Phys.\  {\bf 158} (1984) 142;
  %%CITATION = APNYA,158,142;%%
%\bibitem{Gasser:1984gg}
%  J.~Gasser and H.~Leutwyler,
  %``Chiral Perturbation Theory: Expansions in the Mass of the Strange Quark,''
  Nucl.\ Phys.\ B {\bf 250} (1985) 465.
  %%CITATION = NUPHA,B250,465;%%
%\bibitem{Bijnens:1999sh}
  J.~Bijnens \textit{et. al.}, %G.~Colangelo and G.~Ecker,
  %``The Mesonic chiral Lagrangian of order p**6,''
  JHEP {\bf 9902} (1999) 020,
%  [hep-ph/9902437].
  %%CITATION = HEP-PH/9902437;%%
%\bibitem{Bijnens:2001bb}
%  J.~Bijnens, L.~Girlanda and P.~Talavera,
  %``The Anomalous chiral Lagrangian of order p**6,''
  Eur.\ Phys.\ J.\ C {\bf 23} (2002) 539.
%  [arXiv:hep-ph/0110400].
  %%CITATION = HEP-PH/0110400;%%

\bibitem{Roig:2013baa}
 P.~Roig and J.~J.~Sanz-Cillero,
  %``Consistent high-energy constraints in the anomalous QCD sector,''
  Phys.\ Lett.\ B {\bf 733} (2014) 158.
%  [arXiv:1312.6206 [hep-ph]].
  %%CITATION = ARXIV:1312.6206;%%

 \bibitem{RuizFemenia:2003hm}
  P.~D.~Ruiz-Femen\'{\i}a, A.~Pich and J.~Portol\'es,
  %``Odd intrinsic parity processes within the resonance effective theory of QCD,''
  JHEP {\bf 0307} (2003) 003.
%  [arXiv:hep-ph/0306157].
  %%CITATION = HEP-PH/0306157;%%

\bibitem{Kampf:2011ty}
  K.~Kampf and J.~Novotny,
  %``Resonance saturation in the odd-intrinsic parity sector of low-energy QCD,''
  Phys.\ Rev.\ D {\bf 84} (2011) 014036.
%  [arXiv:1104.3137 [hep-ph]].
  %%CITATION = ARXIV:1104.3137;%%

\bibitem{SDConstr}
%\bibitem{Bijnens:2003rc}
  J.~Bijnens, E.~Gamiz, E.~Lipartia and J.~Prades,
  %``QCD short distance constraints and hadronic approximations,''
  JHEP {\bf 0304} (2003) 055.
%  [hep-ph/0304222].
  %%CITATION = HEP-PH/0304222;%%
%\bibitem{Melnikov:2003xd}
  K.~Melnikov and A.~Vainshtein,
  %``Hadronic light-by-light scattering contribution to the muon anomalous magnetic moment revisited,''
  Phys.\ Rev.\ D {\bf 70} (2004) 113006.
%  [hep-ph/0312226].
  %%CITATION = HEP-PH/0312226;%%
%\bibitem{Ananthanarayan:2004qk}
  B.~Ananthanarayan and B.~Moussallam,
  %``Four-point correlator constraints on electromagnetic chiral parameters and resonance effective Lagrangians,''
  JHEP {\bf 0406} (2004) 047.
%  [hep-ph/0405206].
  %%CITATION = HEP-PH/0405206;%%
%\bibitem{Nyffeler:2009tw}
  A.~Nyffeler,
  %``Hadronic light-by-light scattering in the muon g-2: A New short-distance constraint on pion-exchange,''
  Phys.\ Rev.\ D {\bf 79} (2009) 073012.
%  [arXiv:0901.1172 [hep-ph]].
  %%CITATION = ARXIV:0901.1172;%%
M.~Knecht and A.~Nyffeler, work in progress.

\bibitem{Behrend:1990sr}
  H.~J.~Behrend {\it et al.}  [CELLO Collaboration],
  %``A Measurement of the pi0, eta and eta-prime electromagnetic form-factors,''
  Z.\ Phys.\ C {\bf 49} (1991) 401.
  %%CITATION = ZEPYA,C49,401;%%

\bibitem{Gronberg:1997fj}
  J.~Gronberg {\it et al.}  [CLEO Collaboration],
  %``Measurements of the meson - photon transition form-factors of light pseudoscalar mesons at large momentum transfer,''
  Phys.\ Rev.\ D {\bf 57} (1998) 33.
%  [hep-ex/9707031].
  %%CITATION = HEP-EX/9707031;%%

\bibitem{Aubert:2009mc}
  B.~Aubert {\it et al.}  [BaBar Collaboration],
  %``Measurement of the gamma gamma* ---> pi0 transition form factor,''
  Phys.\ Rev.\ D {\bf 80} (2009) 052002.
%  [arXiv:0905.4778 [hep-ex]].
  %%CITATION = ARXIV:0905.4778;%%

\bibitem{Uehara:2012ag}
  S.~Uehara {\it et al.}  [Belle Collaboration],
  %``Measurement of $\gamma \gamma^* \to \pi^0$ transition form factor at Belle,''
  Phys.\ Rev.\ D {\bf 86} (2012) 092007.
%  [arXiv:1205.3249 [hep-ex]].
  %%CITATION = ARXIV:1205.3249;%%

\bibitem{BABAR:2011ad}
  P.~del Amo Sanchez {\it et al.}  [BaBar Collaboration],
  %``Measurement of the $\gamma \gamma^* --> \eta$ and $\gamma \gamma* --> \eta'$ transition form factors,''
  Phys.\ Rev.\ D {\bf 84} (2011) 052001.
%  [arXiv:1101.1142 [hep-ex]].
  %%CITATION = ARXIV:1101.1142;%%

 \end{thebibliography}



\end{document}